\documentclass[%
 aip,
 amsmath,amssymb,
 reprint,%
]{revtex4-1}

\usepackage{graphicx}
\usepackage{dcolumn}
\usepackage{bm}

\usepackage[utf8]{inputenc}
\usepackage[T1]{fontenc}
\usepackage{mathptmx}
\usepackage{etoolbox}

\makeatletter
\def\@email#1#2{%
 \endgroup
 \patchcmd{\titleblock@produce}
  {\frontmatter@RRAPformat}
  {\frontmatter@RRAPformat{\produce@RRAP{*#1\href{mailto:#2}{#2}}}\frontmatter@RRAPformat}
  {}{}
}%
\makeatother
\begin{document}

\preprint{AIP/123-QED}

\title {Efficient Control of Magnetization Dynamics Via W/CuO$_\text{X}$ Interface}
\author{Antarjami Sahoo}
 \affiliation {Laboratory for Nanomagnetism and Magnetic Materials (LNMM), School of Physical Sciences, National Institute of Science Education and Research (NISER), an OCC of Homi
Bhabha National Institute (HBNI), Jatni 752050, Odisha, India} 
\author{Haifeng Ding}
\affiliation {National Laboratory of Solid State Microstructures, Department of Physics, Nanjing University and Collaborative Innovation Center of Advanced Microstructures, Nanjing 210093,People's Republic of China
}

\author{Antonio Azevedo}
\affiliation {Departamento de Física, Universidade Federal de Pernambuco, Recife, Pernambuco 50670-901, Brazil} 

\author{Subhankar Bedanta}
\email{sbedanta@niser.ac.in}
\affiliation {Laboratory for Nanomagnetism and Magnetic Materials (LNMM), School of Physical Sciences, National Institute of Science Education and Research (NISER), an OCC of Homi
Bhabha National Institute (HBNI), Jatni 752050, Odisha, India}

\date{\today}

\begin{abstract}
Magnetization dynamics, which determine the speed of magnetization switching and spin information propagation, play a central role in modern spintronics. Gaining its control will satisfy the different needs of various spintronic devices. In this work, we demonstrate that the surface oxidized Cu (CuO$_\text{X}$) can be employed for the tunability of magnetization dynamics of ferromagnet (FM)/heavy metal (HM) bilayer system. The capping CuO$_\text{X}$ layer in CoFeB/W/CuO$_\text{X}$ trilayer reduces the magnetic damping value in comparison with the CoFeB/W bilayer. The magnetic damping even becomes lower than that of the CoFeB/CuO$_\text{X}$ by $\sim$16\% inferring the stabilization of anti-damping phenomena. Further, the reduction in damping is accompanied by a very small reduction in the spin pumping-induced output DC voltage in the CoFeB/W/CuO$_\text{X}$ trilayer. The simultaneous observation of anti-damping and spin-to-charge conversion can be attributed to the orbital Rashba effect observed at the HM/CuO$_\text{X}$ interface. Our experimental findings illustrate that the cost-effective CuO$_\text{X}$ can be employed as an integral part of modern spintronics devices owing to its rich underneath spin-orbital physics.
\end{abstract}

\maketitle

\section{\label{sec:level1}Introduction}

Intensive research on spintronics such as magnetic random access memory (MRAM), has been carried out in the last decade, as it shows potential for lower power consumption and instant-on capability \cite{ref1}. Spintronics takes the advantage of magnetic materials which helps in information storage and non-volatile operation. At the same time, the endurance of spintronics devices is far superior to that of resistive and phase-change memory technologies \cite{ref1}. Spintronics also unravels a rich array of fundamental physics arising due to the bulk, interface, and external perturbation induced phenomena \cite{ref2}. The efficient electrical control of magnetization in spin-orbit-torque (SOT) based devices is important for future spintronics applications. The magnetic damping is a critical parameter that determines the required energy and operational speed of SOT-MRAM devices. The scattering of the magnons by the conduction electrons limits the opportunity of reducing the magnetic damping of ferromagnet (FM) metals \cite{Schoen2016}. Several valiant efforts have been put forward to minimize the damping of FMs for future spin-orbitronics applications \cite{Schoen2016}. Ferrimagnetic insulators, like YIG and Heusler alloys, are found experimentally to exhibit low damping. However, the insulating ferrimagnets cannot be a suitable component as some spin-orbitronics applications require the flow of charge current through the magnetic layer. Meantime, the fabrication protocols like high-temperature annealing hinder the integration of Heusler alloys into the CMOS technology \cite{Schoen2016}. Hence, it is highly required to find alternative methodologies for reducing the magnetic damping of FM materials that are compatible with current information technology. 
\par
Recently, the damping of some FM/Nonmagnet (NM) bilayers have been interestingly found to be less compared to that of the parent FMs \cite{behera2016anomalous}, \cite{gupta2021simultaneous}. The reduction in damping, also known as anti-damping, in the Py/Pt bilayer has been attributed to the formation of Rashba-like states at the FM/NM interface. The non-equilibrium spin accumulation at such an interface is believed to generate an anti-damping torque via the inverse Rashba Edelstein effect (IREE). Though here the spin Rashba effect (SRE) plays a vital role, anti-damping phenomena can also be realized due to orbital Rashba effect (ORE). Traditionally, the orbital angular momentum (OAM) was believed not to play any significant role in spintronics applications due to its quenching by crystal field in solids. But in recent times, it has been observed that the OAM mediated orbital torque (OT) can induce magnetization dynamics in the FM layer \cite{An2016}, \cite{An2023}. 
\begin{table*}
\caption{\label{tab:table1}Stacking of different heterostructures with their corresponding nomenclatures.}
\begin{ruledtabular}
\begin{tabular}{cc}
  Stacking & Nomenclature\\ \hline
     Si/SiO$_2$ (300 nm)/CFB (7 nm)/CuO$_\text{X}$ (3 nm) & SF \\ 
        Si/SiO$_2$ (300 nm)/CFB (7 nm)/ W (5 nm) & SA \\ 
        Si/SiO$_2$ (300 nm)/CFB (7 nm)/W (5 nm)/CuO$_\text{X}$ (3 nm) & SA1 \\
        Si/SiO$_2$ (300 nm)/CFB (7 nm)/Cu (10 nm)/W (5 nm) & SB \\
        Si/SiO$_2$ (300 nm)/CFB (7 nm)/Cu (10 nm)/W (5 nm)/CuO$_\text{X}$ (3 nm) & 
 SB1 \\
\end{tabular}
\end{ruledtabular}
\end{table*}
For example, a large current induced SOT is evident in the surface oxidized Cu/Py bilayer even without the presence of any large SOC material \cite{An2016}, \cite{An2023}. This has been attributed to the OAM based OT in CuO$_\text{X}$, which does not require any large SOC materials for its origin. Further, the evolution of ORE/OT can be mainly interfacial, as the resistivity of CuO$_\text{X}$ is expected to be very high. Hence, the ORE, similar to the SRE can be thought of playing a vital role in controlling the magnetization dynamics of adjacent Py layer. The OAM polarization arising due to the orbital hybridization interacts with the electric field generated at the interfaces with structural asymmetry. This leads to the formation of OAM texture of electronic states in the k-space, which is known as the ORE \cite{Go2021}, \cite{Go2017}. Though the ORE has a similarity with the SRE, a high SOC at the interface is not required, unlike SRE for evolution ORE. Upon the application of an external electric field, a non-equilibrium OAM accumulates at the boundaries, and this process is known as orbital Rashba-Edelstein effect (OREE). It is the orbital counterpart of spin Rashba-Edelstein effect and onsager reciprocal effect is known as the inverse orbital Rashba-Edelstein effect (IOREE). The OAM accumulation can modulate the magnetization dynamics of adjacent FM layer with finite SOC via orbital torque. This type of electrically generated torque helps in the development of modern magnetic nanodevices. For example, a large torque has been reported in CoFe/Cu/Al$_2$O$_3$ heterostructures without the requirement of heavy metals (HMs) \cite{kim2021nontrivial}. The extremely large Hall conductivity could not be explained by simple spin torque mechanism and the OAM mediated orbital Edelstein effect at the Cu/Al$_2$O$_3$ interface was attributed to this type of exotic magnetization dynamics phenomena. It has also been found that the CuO$_\text{X}$ capping significantly enhances the SOT efficiency in the thulium iron garnet (TmIG)/Pt bilayer \cite{Ding2020}. Though the CuO$_\text{X}$ has weak SOC, the inversion symmetry breaking at the Pt/CuO$_\text{X}$ interface leads to the orbital current via OREE and hence, the nonlocal generation of SOTs in TmIG/Pt/CuO$_\text{X}$ trilayer \cite{Ding2020}. The spin current injected into the Pt/CuO$_\text{X}$ interface in YIG/Pt/CuO$_\text{X}$ trilayer also gets converted to the charge current with higher efficiency compared to the YIG/Pt via IOREE process \cite{Santos2023}. In both of these examples, the use of heavy metal Pt helps in harnessing OAM via the orbital-to-spin conversion mechanism due to its high SOC. Hence, the combination of HM and surface oxidized Cu can play a vital role in determining the magnetization dynamics of adjacent FMs, which can be detected by either OREE or IOREE. In this manuscript, we report the interesting magnetization dynamics in Co$_{20}$Fe$_{60}$B$_{20}$ (CFB)/W/CuO$_\text{X}$ via ferromagnetic resonance study. The CuO$_\text{X}$ capping reduces the magnetic damping of the trilayer to a value below the damping of CFB without the application of any external DC current, while the spin-to-charge current conversion in CFB/W/CuO$_\text{X}$ trilayer still remains significant. The anti-damping phenomenon can be attributed to the IOREE effect at the W/CuO$_\text{X}$ interface, and it paves an alternative path for lowering the magnetic damping of FMs, and consequently, the fabrication of power-efficient spintronics devices.    
\section{\label{sec:level2}Experimental Methods}

\begin{figure*}
\includegraphics[scale=0.55]{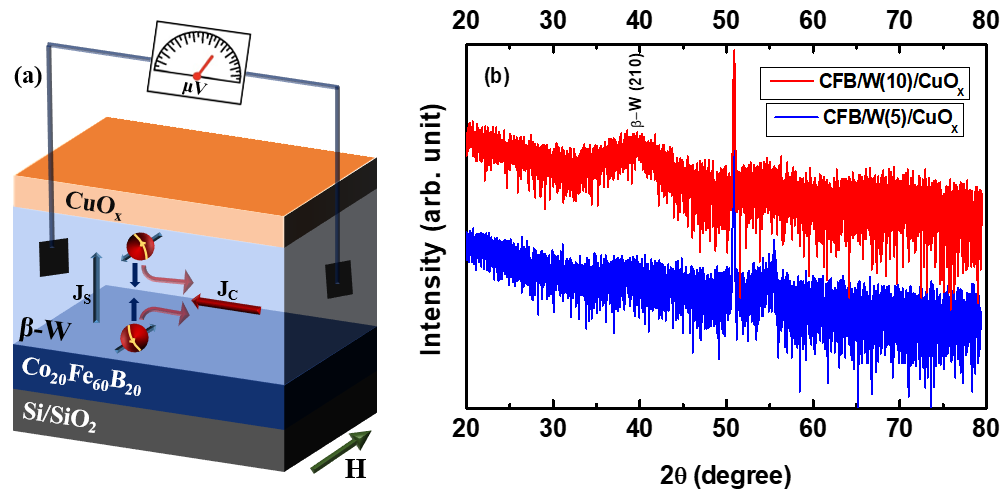}
\caption{\label{fig:1} (a) Schematic of Si/SiO$_2$/CFB/W/CuOx heterostructure illustrating the spin to charge conversion phenomena, (b) GIXRD patterns of Si/SiO$_2$/CFB (7 nm)/W (5, 10 nm)/CuOx (3 nm) heterostructures}
\end{figure*}

Four different types of heterostructures with CFB (7 nm)/Cu (0, 10 nm)/W (5 nm)/CuO$_\text{X}$ (0, 3 nm) and CFB (7 nm)/CuO$_\text{X}$ (3 nm) stacking have been fabricated on Si/SiO$_2$ (300 nm) substrates for the investigation of magnetization dynamics and spin pumping phenomena. In addition, the CFB (7 nm)/W (10 nm)/CuO$_\text{X}$ (3 nm) heterostructure was also fabricated to reaffirm the stabilization of $\beta$ phase of W. The heterostructures can be read as shown in TABLE \ref{tab:table1}.
The CFB, $\beta$-W, and Cu layers were grown by DC magnetron sputtering (Manufactured by EXCEL Instruments, India) at room temperature. The top thin Cu layer oxidizes naturally to form the CuO$_\text{X}$ capping in SF, SA1 and SB1 heterostructures. We have also fabricated a 10 nm Cu thin film on Si/SiO$_2$ to investigate the formation of the natural oxidation of Cu. Before the fabrication of heterostructures, thin films of CFB, W, and Cu were prepared for thickness calibration and study of magnetic and electrical properties. The base pressure of the sputtering chamber was usually maintained at $\sim$ 4×10$^{-8}$ mbar prior to the deposition. The structural characterizations of individual thin films and heterostructures were performed by x-ray diffraction (XRD) and x-ray reflectivity (XRR) measurements. Before the fabrication of heterostructures, the thickness of individual layers was calibrated via XRR, and the growth of different materials of the heterostructures was monitored using a quartz crystal monitor. The superconducting quantum interference device based vibrating sample magnetometer (SQUID-VSM) was employed for the static magnetization characterization. The magnetization dynamics were investigated by a lock-in based ferromagnetic resonance (FMR) spectrometer manufactured by NanOsc, Sweden. The heterostructures were kept in a flip-chip manner on the co-planner waveguide (CPW) and the FMR spectra in 4-17 GHz range were recorded for all the samples. The FMR spectrometer set-up is also equipped with an additional nano-voltmeter using which spin-to-charge conversion phenomena of all the devices are measured via Inverse Spin Hall Effect (ISHE) with 15 dBm RF power. The contacts were given at the two opposite ends of 3 mm × 2 mm devices using silver paste to measure the spin pumping induced voltage difference across the samples \cite{Singh2018}. The schematics illustrating the CFB/W/CuO$_\text{X}$ heterostructure and spin-to-charge conversion process owing to the spin injection from CFB are shown in Fig. \ref{fig:1} (a). 

\section{\label{sec:level3}Results and Discussion}

The grazing incidence x-ray diffraction (GIXRD) was performed for all the heterostructures. The XRD patterns of CFB (7 nm)/W (5, 10 nm)/CuO$_\text{X}$ (3 nm) heterostructures are shown in Fig. 1 (b). The presence of (210) Bragg’s peaks of W at $\sim$ 39.8$^\circ$ indicates the stabilization of $\beta$ phase of W (A-15 crystal structure) \cite{hait2022spin},\cite{behera2017capping}. The Bragg’s peaks are more prominent for heterostructures with 10 nm thick W layers as diffraction intensity increases with increasing W thickness. The XRD patterns for SA, SA1, SB, and SB1 are similar indicating W is of the $\beta$-phase in all these heterostructures. Here, we have not used reactive gases like O$_2$ and N$_2$ for the growth of $\beta$-W unlike some previous reports \cite{mchugh2020impact}. We don’t observe other Bragg’s peaks of W in the XRD patterns, inferring an oriented growth of $\beta$-W. We do not also observe the (110), (200), and (210) Bragg’s peaks for the bcc $\alpha$-W in the XRD patterns \cite{behera2017capping}. Further, the fullwidth half maximum of (210) Bragg’s peaks of W decreases as we increase the thickness of W (Fig. S1). The stabilization of this type of oriented $\beta$ phase of W is quite important for future SOT device fabrication, as $\beta$ phase of W yields high spin-orbit coupling.    

\begin{figure*}
\includegraphics[scale=0.45]{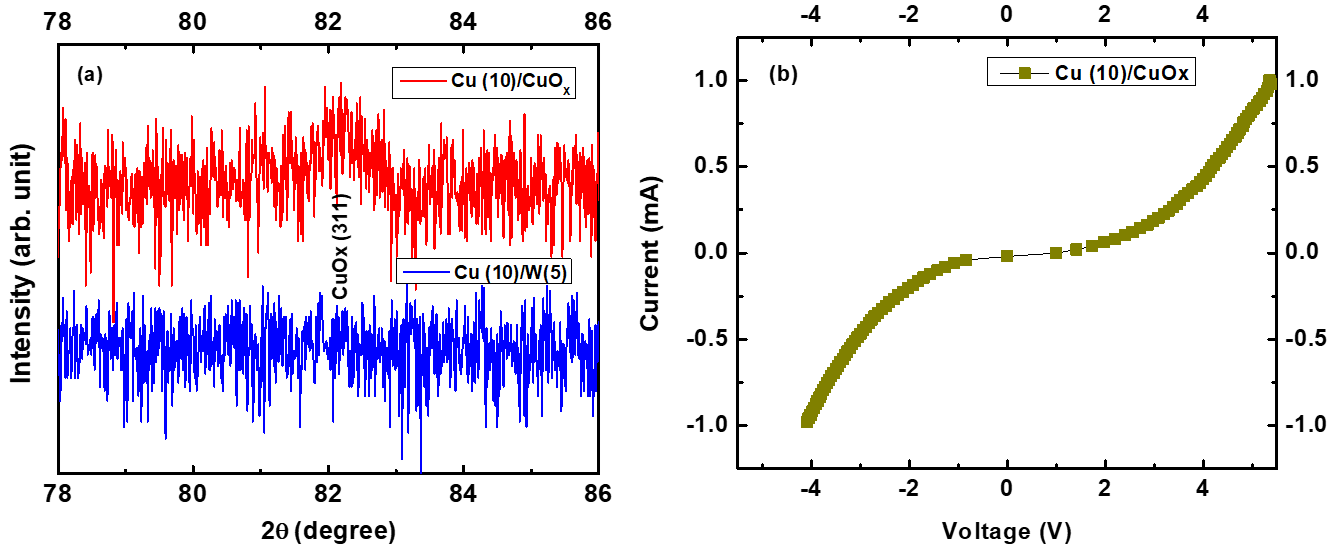}
\caption{\label{fig:2} (a) GIXRD patterns of Si/SiO$_2$/Cu (10 nm) and Si/SiO$_2$/CFB (7 nm)/Cu (10 nm)/W (5 nm) heterostructures, (b) I-V plot of Si/SiO$_2$/Cu (10 nm) heterostructure}
\end{figure*}

\begin{figure*}
\includegraphics[scale=0.45]{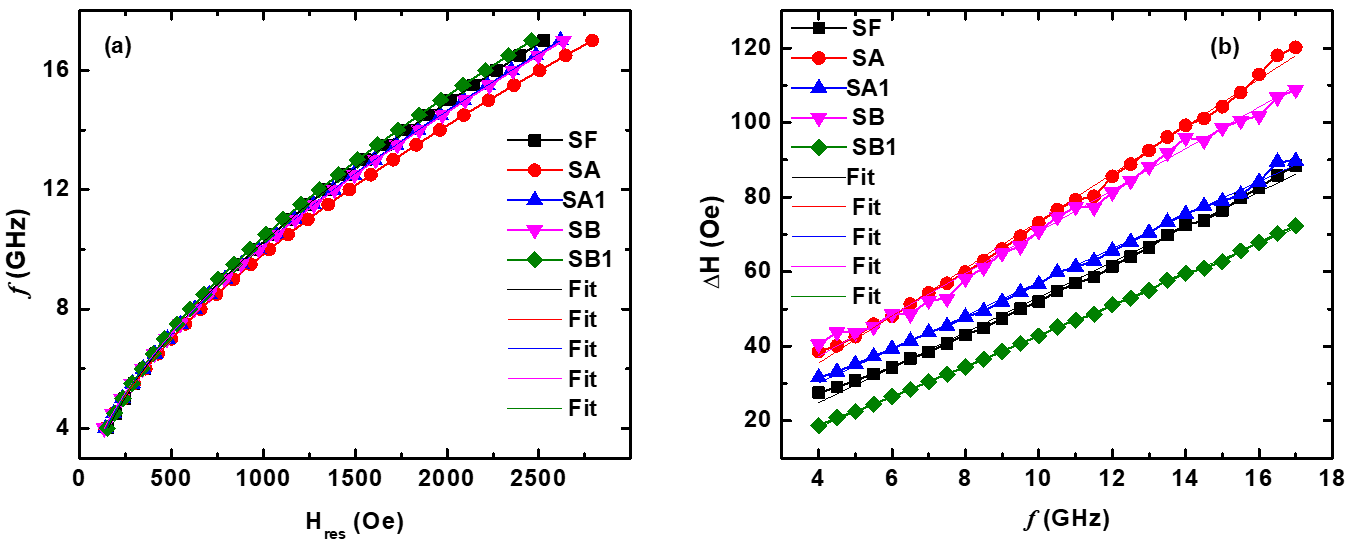}
\caption{\label{fig:3} (a) Frequency ($f$) versus resonance field ($H_{res}$) and (b) linewidth ($\Delta H$) versus frequency ($f$) behaviour for all the heterostructures listed in TABLE \ref{tab:table1}. The solid lines in (a) and (b) are the best fits to equations 2 and 3, respectively.}
\end{figure*}
The Cu capping layer was allowed to oxidize naturally in ambient environment at laboratory. After that, the formation of the top CuO$_\text{X}$ layer in SA1 and SB1 heterostructures was confirmed by GIXRD and I-V measurements. The 10 nm thick Cu layer without any capping displays a Bragg’s peak at ~ 82.2$^\circ$, whereas the same peak is absent for the SB heterostructure, where the 10 nm Cu is capped with 5 nm W (Fig. 2 (a)). The Bragg’s peak at $\sim$ 82.2$^\circ$ has also been previously reported for Cu film \cite{AlSaeedi2021}, inferring the formation of surface oxidized Cu in the SA1 and SB1 heterostructures. Further, the I-V measurement was also performed for the bare 10 nm thick Cu film. The I-V plot (Fig. 2 (b)) shows a non-linear behavior, and it is quite similar to the I-V properties of the semiconducting copper oxide films \cite{singh2014relationship}, \cite{singh2011electronic}. These experiments reaffirm the natural oxidation of Cu if not capped by any other layer. In fact, a similar natural oxidation of Cu has also been reported previously and has been attributed to the evolution of IOREE \cite{Santos2023}.
\par
Next, we study the effect of the top CuO$_\text{X}$ on the magnetization dynamics of heterostructures via FMR. The heterostructures are placed in a flip-chip manner on CPW. Fig. S2 shows the typical FMR spectra of SA1 measured in the 4-17 GHz range. Similar types of FMR spectra were also recorded for other heterostructures. All the FMR spectra were fitted to the derivative of symmetric and antisymmetric Lorentzian function to evaluate the resonance field ($H_{res}$) and full-width-at-half-maximum of magnetic field swept-absorption ($\Delta H$) \cite{Singh2018},\cite{belmeguenai2018ferromagnetic}: 
\begin{widetext}
\begin{equation}
    \text{FMR Signal }= K_1 \frac{4(\Delta H)(H-H_{res})}{[(\Delta H)^2 + 4 (H - H_{res})^2]^2} - K_2 \frac{(\Delta H)^2 - 4(H-H_{res})^2}{[(\Delta H)^2 + 4 (H - H_{res})^2]^2} + \text{Offset},
    \label{eq:FMR}
\end{equation}
\end{widetext}
where $K_1$ and $K_2$ are the antisymmetric and symmetric absorption coefficients, respectively.
The $H_{res}$ and $\Delta H$ extracted for various resonance frequencies from the Lorentzian fit of the field dependent FMR absorption are shown in Fig. 3. The $H_{res}$ dependent $f$ of different heterostructures are plotted in Fig. 3 (a). The $f$ vs $H_{res}$ plots are fitted using the Kittel’s equation \cite{Singh2018},\cite{belmeguenai2018ferromagnetic}:
\begin{equation}
    f = \frac{\gamma}{2 \pi} \sqrt{(H_K + H_{res})(H_K +H_{res}+4\pi M_{eff})},
    \label{eq:Kittel}
\end{equation}
where$$4\pi M_{eff} = 4\pi M_S + \frac{2K_S}{M_S t_{FM}}$$
and $H_K$, $K_S$, and $t_{FM}$ are the anisotropy field, perpendicular surface anisotropy constant, and the thickness of FM, respectively. Further, $\gamma$ is the gyromagnetic ratio and $4\pi M_{eff}$ represents the effective magnetization. The $4 \pi M_{eff}$ extracted from the fitting gives similar values as compared with the saturation magnetization value ($4 \pi M_{S}$) calculated from the SQUID VSM measurements. Further, the effective Gilbert damping constant and hence, the magnetization relaxation mechanism are studied from the resonance frequency dependent FMR linewidth behavior. The frequency domain measurement of ferromagnetic resonance linewidth $\Delta H$ allows the separation of intrinsic and extrinsic contributions to the magnetic damping by the following equation, 
\begin{equation}
    \Delta H = \Delta H_0 + \frac{4 \pi \alpha_{eff}}{\gamma}f,
    \label{eq:damping}
\end{equation}
where the $\Delta H_0$ is known as the inhomogeneous linewidth and represents the extrinsic contribution to the damping of the precessing magnetization. The value of $\alpha_{eff}$ represents the intrinsic contribution to the damping.  The $\Delta H$ vs $f$  plots of all the heterostructures are shown in Fig. 3 (b). All the resonance frequency dependent $\Delta H$ plots are fitted by the equation 3 to evaluate the effective Gilbert damping ($\alpha_{eff}$) constant \cite{belmeguenai2018ferromagnetic}. The linear dependency of $\Delta H$ on $f$ indicates the magnetic damping is mainly governed by intrinsic mechanism via electron-magnon scattering rather than the extrinsic two magnon scattering. The energy during the magnetization precession can also be transferred between the uniform and non-uniform precession modes via two-magnon scattering, resulting an additional contribution to the intrinsic damping in $\alpha_{eff}$. This usually leads to non-linear frequency-dependent $\Delta H$ behavior. As we do not observe the non-linearity in the $\Delta H$ vs $f$ plots, the contribution of the two-magnon scattering can be neglected.

\begin{figure*}
\includegraphics[scale=0.5]{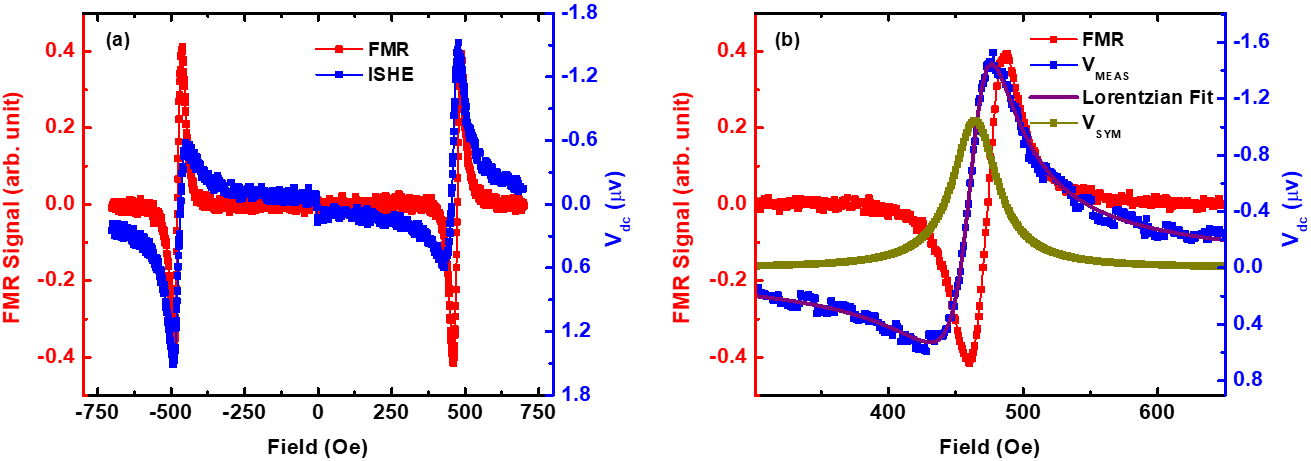}
\caption{\label{fig:4} (a) The magnetic field swept FMR spectra and ISHE pattern of SA1 (Si/SiO$_2$/CFB (7 nm)/W (5 nm)/CuOx (3 nm)) heterostructure; (b) FMR Spectra, $V_{MEAS}$, Lorentzian Fit to $V_{MEAS}$ and $V_{SYM}$ versus applied magnetic field of SA1 (Si/SiO$_2$/CFB (7 nm)/W (5 nm)/CuOx (3 nm)) heterostructure.}
\end{figure*}

\begin{table*}
\caption{\label{tab:table2} Effective magnetic damping of various heterostructures.}
\begin{ruledtabular}
\begin{tabular}{cccccc}
   &  SF &  SA &  SA1 &  SB & SB1 \\ \hline
  $\alpha_{eff}$ ($\pm$ 0.0001) & 0.0060 & 0.0070 & 0.0053 & 0.0065 & 0.0051 \\
\end{tabular}
\end{ruledtabular}
\end{table*}

The values of $\alpha_{eff}$ of different heterostructures are shown in TABLE \ref{tab:table2}. The $\alpha_{eff}$ of CFB/W is found to be larger compared to that with CFB/CuO$_\text{X}$, inferring possible spin-pumping due to the presence of a high SOC $\beta$-W layer. Interestingly, the magnetic damping of CFB/W/CuO$_\text{X}$ decreases and even becomes less than that of CFB/CuO$_\text{X}$. Such type of decrease can be termed as anti-damping, which has been established in the heterostructures without flowing any DC current. To further confirm the anti-damping behaviour, the dynamics of another pair of heterostructures, SB and SB1, where a thick Cu layer is inserted between CFB and W, have also been investigated. The SB heterostructure, where there is no top CuO$_\text{X}$ capping, shows an enhancement in damping similar to SA (TABLE \ref{tab:table2}). While the $\alpha_{eff}$ in SB1, where the CFB/Cu/W is capped with CuO$_\text{X}$ again becomes lower compared to that of the CFB/CuO$_\text{X}$. Hence, we can conclude that the W/CuO$_\text{X}$ interface plays a vital role in controlling the magnetization precession. The naturally oxidized Cu layer has proven to exhibit OREE, both experimentally and theoretically \cite{Go2021},\cite{Go2017},\cite{kim2021nontrivial},\cite{Ding2020}. Thus, we can expect the evolution of orbital Rashba state at the W/CuO$_\text{X}$ interface. The high SOC of W can facilitate the spin-to-orbital conversion of angular momentum, which can lead to the accumulation of OAM at the W/CuO$_\text{X}$ interface. The accumulated OAM can be converted to the charge current via IOREE, which can induce anti-damping like torque similar to the anti-damping torque experienced at Py/Pt interface due to the non-equilibrium spin accumulation \cite{behera2016anomalous}, \cite{behera2017capping}. As the spin diffusion length of Cu is quite large, the 10 nm-thick Cu spacer layer efficiently transports the spin current from CFB to W and vice-versa. Hence, the SB1 heterostructure also exhibits the anti-damping behaviour similar to SA1 arising due to W/CuO$_\text{X}$ interface. The lowering of magnetic damping value by the HM/CuO$_\text{X}$ overlayer can be used as an effective tool for achieving low damping magnetic materials, which is important for power-efficient spintronics applications. 
\par
As the W/CuO$_\text{X}$ interface induces anti-damping, it is also important to investigate the spin-to-charge conversion in these types of heterostructures. The spin pumping induced charge current measurements were performed for all the heterostructures. Fig. 4 (a) shows the typical field-dependent DC voltage (V$_{dc}$) measured across the SA1 heterostructure under FMR conditions. The measured DC voltage reverses its polarity when the magnetic field reverses its direction. This confirms the presence of spin pumping-induced spin-to-charge conversion in our heterostructures. A similar type of ISHE has also been observed in SA. As the presence of thick Cu layer significantly reduces the resistance of the devices, accurate spin pumping-induced charge current measurement of SB and SB1 heterostructures was not possible. The lower resistances of the SB and SB1 samples produce voltages in the tens of nV and hence, the noise level plays a vital role in the measured data. In order to separate the symmetric (V$_{SYM}$) and asymmetric (V$_{ASYM}$) components, the V$_{MEAS}$ vs $H$ plots were fitted with the following Lorentzian function:

\begin{figure*}
\includegraphics[scale=0.55]{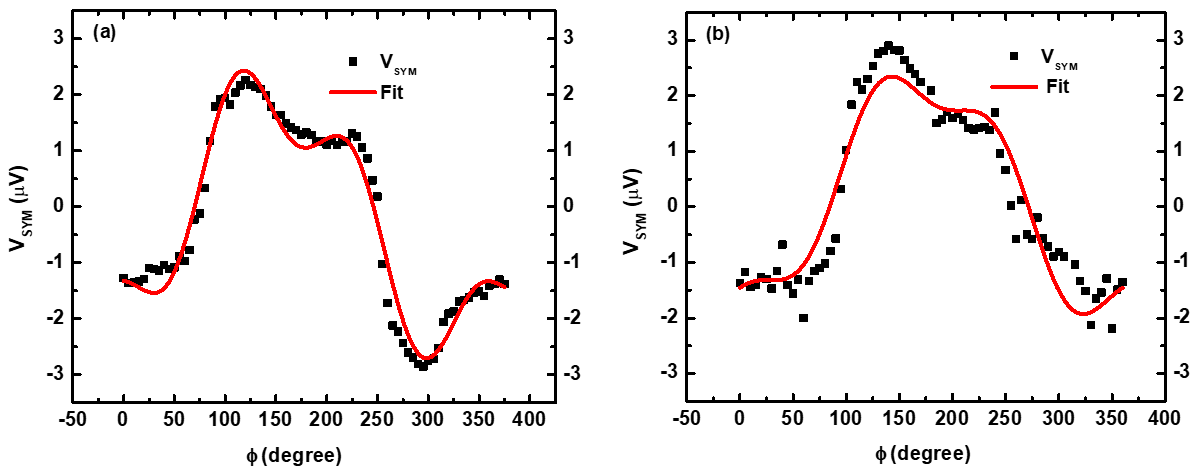}
\caption{\label{fig:2} Angel-dependent V$_{SYM}$ and their fits for (a) CFB (7 nm)/W (5 nm) [SA] and (b) CFB (7 nm)/W (5 nm)/CuO$_\text{X}$ (3 nm) [SA1] heterostructures}
\end{figure*}  
\begin{widetext}

\begin{equation}
    V_{dc} = V_{SYM} \frac{(\Delta H)^2}{(\Delta H)^2 + (H - H_{res})^2} + V_{ASYM} \frac{(\Delta H) (H-H_{res})}{(\Delta H)^2 + (H - H_{res})^2} 
    \label{eq:ISHE}
\end{equation}
\end{widetext}
The V$_{MEAS}$, its Lorentzian fit, and V$_{SYM}$ obtained around the resonance condition of SA1 are shown in Fig. 4 (b).

\par
The angle-dependent spin-to-charge conversion measurements were performed for both the SA and SA1 heterostructures. The V$_{SYM}$ vs $\phi$ data (Fig. 5) were fitted with the following equation to exclude the spin rectification effects and evaluate the spin pumping voltage (V$_{SP}$) \cite{Singh2018}: 
\begin{equation}
\begin{split}
    V_{SYM} = V_{SP} Cos^3 (\phi) +  V_{AHE} Cos(\phi) Cos(\theta) \\ + V_{SYM}^{AMR\perp} Cos (2\phi) Cos(\phi) + V_{SYM}^{AMR\parallel} Sin (2\phi) Cos(\phi)
\end{split}
\end{equation}
The V$_{SP}$ for SA and SA1 heterostructures was found to be $\sim$ 5.6 µV and $\sim$ 4.9 µV, respectively. The small reduction of V$_{SP}$ of CFB/W/CuO$_\text{X}$ heterostructure compared to CFB/W bilayer can be due to IOREE induced spin-to-charge conversion current at the W/CuO$_\text{X}$ interface, which is in opposite direction to the already present ISHE-induced charge current in CFB/W. Here, We cannot also completely neglect the effect of slight variation in W thickness, which can induce a small change in V$_{SP}$. However, the anti-damping phenomena, which mainly corroborate the spin accumulation at W/CuO$_\text{X}$ interface, infer the IOREE mechanism to be the most prominent one.The magnetic damping in FM/HM bilayer usually increases when the spin pumping-induced spin-to-charge conversion is present in the system. This can also be seen in our CFB/W bilayer. Whereas, interestingly, the presence of the top CuO$_\text{X}$ capping layer reduces the magnetic damping without much affecting the spin-to-charge conversion current. Thus, this type of system, where the reduction in damping and efficient spin-to-charge interconversion are observed simultaneously, can bring about a paradigm shift in spintronics device applications. It can certainly pave the path for the development of low-operating-power spintronics devices.

\section{Conclusion}
In conclusion, we have presented the control of magnetic damping of FM/HM bilayer when capped with surface oxidized Cu. The CuO$_\text{X}$ layer lowers the damping value of CoFeB/W/CuO$_\text{X}$ trilayer even below that for the CoFeB/CuO$_\text{X}$. The strong anti-damping via CuO$_\text{X}$ capping is accompanied by a very small reduction in the spin-to-charge interconversion phenomena as evident in the spin pumping-FMR experiment. The rich spin-orbital physics leading to the inverse orbital Rashba effect at HM/CuO$_\text{X}$ interface can attribute to the simultaneous observation of anti-damping and sizable spin-pumping induced output voltage.
\begin{acknowledgments}
We acknowledge the Department of Atomic Energy (DAE), the Department of Science and Technology (DST) of the Government of India, SERB project CRG/2021/001245. A.S. acknowledges the DST-National Postdoctoral Fellowship in Nano Science and Technology. 
\end{acknowledgments}
\section*{Reference}
\nocite{*}
\bibliography{aipsamp}

\end{document}